\documentstyle[prl,aps,multicol]{revtex}
\begin{document}

\draft

\title{Initial Stages of Bose-Einstein Condensation}

\author{H.T.C. Stoof}
\address{Institute for Theoretical Physics,
         University of Utrecht, Princetonplein 5, \\
         P.O. Box 80.006, 3508 TA  Utrecht,
         The Netherlands} 

\maketitle

\begin{abstract}
We present the quantum theory for the nucleation of Bose-Einstein condensation in a dilute atomic Bose gas. This quantum theory comfirms the results of the semiclassical treatment, but has the important advantage that both the kinetic and coherent stages of the nucleation process can now be described in a unified way by a single Fokker-Planck equation.    
\end{abstract}


\begin{multicols}{2}
In two previous papers we have developed the semiclassical theory for the nucleation of Bose-Einstein condensation in a weakly-interacting Bose gas \cite{henk}. At that time the main reason for studying this nucleation problem, was to determine whether and on what time scale Bose-Einstein condensation could be observed in ongoing experiments with magneticaly trapped atomic hydrogen \cite{greytak,walraven}, and for that purpose the semiclassical theory was sufficient. However, now that Bose-Einstein condensation has actually been achieved in atomic $^{87}$Rb \cite{eric}, $^{23}$Na \cite{wolfgang} and $^7$Li \cite{randy} vapours, it appears that experimental studies of the dynamics of Bose-Einstein condensation can be performed in such detail, that a more elaborate theory is required to fully understand the outcome of these experiments. It is the aim of this paper to present such a theory. 

In the semiclassical picture the formation of the condensate proceeds in three stages. In the first stage the gas is quenched into the critical region of the phase transition, for example by means of evaporative cooling. This quenching of the gas is a kinetic process and can be accurately described by the quantum Boltzmann equation. However, such a kinetic equation cannot lead to a macroscopic occupation of the one-particle ground state. To achieve that, a second stage is needed in which the gas first develops the instability towards Bose-Einstein condensation and then coherently populates the ground state by a depletion of the low-lying excited states. After this coherent stage the gas has clearly a highly non-equilibrium energy distribution, and must come to equilibrium in a third and final stage. This last stage is again of a kinetic nature and can be studied by the appropriate quantum Boltzmann equation for the Bogoliubov quasiparticles of the Bose condensed gas. 

Although the semiclassical theory gives important insight into the physics of the nucleation of Bose-Einstein condensation and the various time scales involved, it is not completely satisfying because of the following two reasons. First, it makes a sharp distinction between kinetic and coherent processes which in principle of course take place simultaneously. Second, the coherent stage of the evolution is described by a time-dependent nonlinear Schr\"odinger equation for the complex order parameter 
$\phi(\vec{x},t) \equiv \langle \psi(\vec{x},t) \rangle$, 
that formally has the property that if the order parameter is zero initially it will always remain zero. Therefore, the semiclassical theory actually makes use of the fact that there are quantum fluctuations in $\phi(\vec{x},t)$ without taking these explicitly into account. Both these fundamental problems are resolved in the quantum theory, to which we now turn our attention.

To arrive at the quantum theory of the order parameter $\phi(\vec{x},t)$ it is, in contrast to the semiclassical approach, not sufficient to derive the equation of motion for the expectation value of the (Heisenberg) operator $\psi(\vec{x},t)$ that annihilates an atom at position $\vec{x}$ and at time $t$. Instead we need to find an equation that determines the evolution of the full probability distribution $P[\phi^*,\phi;t]$, so that we are able to consider also the fluctuations in $\phi(\vec{x},t)$. This is most easily achieved as follows. Introducing the initial density matrix $\rho(t_0)$ of the at that time still uncondensed gas and the coherent states
\begin{equation}
|\phi(\vec{x}),t \rangle = 
   \exp \left\{
          \int d\vec{x}~ \phi(\vec{x}) \psi^{\dagger}(\vec{x},t)
        \right\} |0 \rangle~,  
\end{equation}
where $|0 \rangle$ denotes the vacuum state, the probability distribution $P[\phi^*,\phi;t]$ of interest equals                                      \begin{equation}
P[\phi^*,\phi;t] = {\rm Tr}
    \left[ \rho(t_0) \frac{|\phi,t \rangle \langle \phi,t|}
                          {\langle \phi,t| \phi,t \rangle}
    \right]~.
\end{equation}
Moreover, using an expansion of $\rho(t_0)$ in terms of the above coherent states, the latter can be rewritten as the functional integral
\begin{equation}
\label{P}
P[\phi^*,\phi;t] = 
  \int d[\phi_0^*]d[\phi_0]~
           \rho[|\phi_0|^2;t_0] 
              \frac{|\langle \phi,t|\phi_0,t_0 \rangle|^2}
                   {\langle \phi,t| \phi,t \rangle}~.
\end{equation}
Since 
$\langle \phi,t| \phi,t \rangle =
    \exp \left\{ \int d\vec{x}~ |\phi(\vec{x})|^2 \right\}$ is in fact independent of time, our task is essentially reduced to the determination of the probability 
$|\langle \phi,t|\phi_0,t_0 \rangle|^2$.   

This we achieve by writing $\langle \phi,t|\phi_0,t_0 \rangle$ as a `path' integral over all field evolutions from $t_0$ to $t$. Similarly, we also write 
$\langle \phi,t|\phi_0,t_0 \rangle^* =
                             \langle \phi_0,t_0|\phi,t \rangle$
as a `path' integral over fields, but these can now be seen as evolving backwards in time from $t$ to $t_0$. Performing then also the integrations over $\phi_0(\vec{x})$ in Eq.~(\ref{P}), we see that we are lead in a natural way to the Keldysh formalism developed in Ref.\ \cite{henk}, because we have to perform a functional integral over all field configurations $\psi(\vec{x},t)$ that evolve from $t_0$ to $t$ and back to $t_0$, i.e.\ along the Keldysh contour $C$. Using the results obtained there, we conclude that the probability distribution for $\phi(\vec{x})$ is given by
\begin{equation}
\label{integral}
P[\phi^*,\phi;t] = 
  \int d[\psi^*]d[\psi]~
    \exp \left\{\frac{i}{\hbar} S[\psi^*,\psi] \right\}~,
\end{equation}
where the effective action $S[\psi^*,\psi]$ in the integrand becomes
\end{multicols}
\begin{eqnarray}
S[\psi^*,\psi] =
  \sum_{\vec{k}} \mbox{\Large\{} 
                    \phi^*_{\vec{k}} \psi_{\vec{k}}(t)
                      + \psi^*_{\vec{k}}(t) \phi_{\vec{k}}
                     &-& |\phi_{\vec{k}}|^2 \mbox{\Large\}} 
+ \int_C dt \int_C dt'~
    \left[ \sum_{\vec{k}}~ \psi^*_{\vec{k}}(t) 
      \left\{ \left( i\hbar \frac{\partial}{\partial t} 
                    - \xi(\vec{k}) \right) \delta(t,t')
              - \hbar \Sigma(\vec{k};t,t') \right\} 
                                          \psi_{\vec{k}}(t')
                   \right.  \nonumber \\
               &-& \left. 
 \frac{1}{2V} \sum_{\vec{k},\vec{k}',\vec{K}}
        V(\vec{k},\vec{k}',\vec{K};t,t')
           \psi^*_{\vec{K}/2+\vec{k}}(t)
           \psi^*_{\vec{K}/2-\vec{k}}(t)
           \psi_{\vec{K}/2-\vec{k}'}(t')
           \psi_{\vec{K}/2+\vec{k}'}(t')
                   \right]~,   
\end{eqnarray}
introducing the notation 
$\xi(\vec{k}) = \hbar^2\vec{k}^2/2m - \mu_0 
              \equiv \epsilon(\vec{k}) - \mu_0$ for the kinetic energy of an atom with momentum $\hbar\vec{k}$ and mass $m$ relative to the chemical potential $\mu_0$ at $t_0$, $\delta(t,t')$ for the $\delta$-function on the Keldysh contour defined by $\int_C dt'~\delta(t,t') = 1$, and $V$ for the volume of the system.    

We arrive at this result by making use of the so-called ladder or many-body T-matrix approximation. Physically, this means that we include all two-body scattering processes in the gas but neglect three (or more) particle processes. This is justified for a weakly-interacting Bose gas because of the smallness of the gas parameter $na^3$, where $n$ is the density and $a$ the two-body scattering length of the atoms in the gas. Furthermore, in the same approximation the selfenergy obeys the Hartree-Fock like expression
\begin{equation}
\hbar\Sigma(\vec{k};t,t') = 2i \int \frac{d\vec{k}'}{(2\pi)^3}~
  V(\vec{k}-\vec{k}',\vec{k}-\vec{k}',\vec{k}+\vec{k}';t,t')
                                            G(\vec{k}';t',t)~,
\end{equation} 
and the effective interaction between the atoms is 
$V(\vec{k},\vec{k}',\vec{K};t,t') =  
       ( T(\vec{k},\vec{k}',\vec{K};t,t') +  
                   T(-\vec{k},\vec{k}',\vec{K};t,t') )/2$
in terms of the many-body T-matrix that is determined by the Lippman-Schwinger equation
\begin{eqnarray}
\label{LS}
T(\vec{k},\vec{k}',\vec{K};t,t') &=&           
  V(\vec{k}-\vec{k}') \delta(t,t')  \nonumber \\ 
&+& \frac{i}{\hbar} \int_C dt'' \int \frac{d\vec{k}''}{(2\pi)^3}~
       V(\vec{k}-\vec{k}'')
            G(\vec{K}/2+\vec{k}'';t,t'')
            G(\vec{K}/2-\vec{k}'';t,t'')
                T(\vec{k}'',\vec{k}',\vec{K};t'',t')~,
\end{eqnarray}
\vspace*{-0.2in}
\begin{multicols}{2}
\noindent
with $V(\vec{k}-\vec{k}')$ the Fourier transform of the interatomic potential, 
\begin{equation}
G(\vec{x},t;\vec{x}',t') \equiv 
   -i~ {\rm Tr} \left[ \rho(t_0) T_C 
               \left( \psi(\vec{x},t) 
                      \psi^{\dagger}(\vec{x}',t') \right)
               \right]
\end{equation}
the one-particle Green's function and $T_C$ the time-ordening operator along the Keldysh contour.

To extract the equation of motion for $P[\phi^*,\phi;t]$ from Eq.~(\ref{integral}), we want to express the functional integral over the fields $\psi(\vec{x},t_+)$ and $\psi(\vec{x},t_-)$, which respectively live on the forward and backward branch of the Keldysh contour, as a functional integral over fields that are defined on the real axis. This is convenient, because in this manner we can from the result immediately read off the appropriate `Schr\"odinger' equation for the probability distribution. We achieve this goal in the following way: First, we perform the variable transformation 
$\psi(\vec{x},t_\pm) = \phi(\vec{x},t) \pm \chi(\vec{x},t)/2$
in Eq.~(\ref{integral}). Substituting this transformation into the effective action $S[\psi^*,\psi]$, it turns out to be convenient to decompose the various functions $F(t,t')$ on the Keldysh contour into their retarded, advanced and Keldysh components $F^{(\pm)}(t,t')$ and $F^K(t,t')$, respectively \cite{keldysh}. Second, we notice that for the ultracold atomic gases of interest the thermal de Broglie wavelength 
$\Lambda = (2\pi\hbar^2/mk_BT)^{1/2}$ is so large compared to the scattering length $a$ that we are allowed to make a Markovian approximation to the many-body T-matix and hence to the selfenergy $\hbar\Sigma(\vec{k};t,t')$, i.e.\ we can use 
$F^{(\pm),K}(t,t') \simeq F^{(\pm),K}(t) \delta(t-t')$ for these quantities. Third, we expand the action up to quadratic order in the field $\chi(\vec{x},t)$, which physically describes the quantum fluctuations in the order parameter 
$\langle \psi(\vec{x},t) \rangle$ that are neglected in the semiclassical theory, and then integrate out this field. It is important to note that the cubic terms in the action should indeed be neglected here to avoid a double counting of the effects of the interaction \cite{henk}. 
  
After performing the integration over $\chi(\vec{x},t)$ we clearly arrive at an expression for $P[\phi^*,\phi;t]$ as a functional integral over $\phi(\vec{x},t)$. Using for example the methods of canonical quantization, we are now in a position to derive the `Schr\"odinger' equation that is solved by this functional integral. Carrying out the above procedure requires some straightforward but tedious algebra and ultimately leads to the central result of the present paper, i.e.\ a Fokker-Planck equation describing both the kinetic and the coherent stages of the condensation process. It reads \cite{phdif}
\end{multicols}
\begin{eqnarray} 
\label{FP} 
i\hbar \frac{\partial}{\partial t} P =
 - \sum_{\vec{k}} 
   \frac{\partial}{\partial \phi_{\vec{k}}}
       \left( \xi(\vec{k})\phi_{\vec{k}} 
            + \hbar\Sigma^{(+)}(\vec{k};t)\phi_{\vec{k}}
            + \frac{1}{V} \sum_{\vec{k}',\vec{k}''} 
               V^{(+)}(\vec{k},\vec{k}',\vec{k}'')
                   \phi^*_{\vec{k}'+ \vec{k}''-\vec{k}}  
                   \phi_{\vec{k}''} \phi_{\vec{k}'}
       \right) P                                 \nonumber \\
+ \sum_{\vec{k}}
  \frac{\partial}{\partial \phi^*_{\vec{k}}}
       \left( \xi(\vec{k})\phi^*_{\vec{k}} 
            + \hbar\Sigma^{(-)}(\vec{k};t)\phi^*_{\vec{k}}
            + \frac{1}{V} \sum_{\vec{k}',\vec{k}''}
               V^{(-)}(\vec{k},\vec{k}',\vec{k}'')
                   \phi^*_{\vec{k}'} \phi^*_{\vec{k}''}
                   \phi_{\vec{k}'+ \vec{k}''-\vec{k}}
       \right) P                                \nonumber \\
- \frac{1}{2} \sum_{\vec{k},\vec{k}'} 
  \frac{\partial^2}{\partial \phi^*_{\vec{k}}
                    \partial \phi_{\vec{k}'}}
       \left( \hbar\Sigma^K(\vec{k};t) \delta_{\vec{k},\vec{k}'}
            + \frac{2}{V} \sum_{\vec{k}''} 
               V^K(\vec{k},\vec{k}',\vec{k}'')
                   \phi^*_{\vec{k}'+ \vec{k}''-\vec{k}}
                   \phi_{\vec{k}''}
       \right) P~,
\end{eqnarray}
\vspace*{-0.2in}
\begin{multicols}{2}
\noindent
where $V^{(\pm),K}(\vec{k},\vec{k}',\vec{k}'')$ is a shorthand notation for 
$V^{(\pm),K}(\vec{k}-(\vec{k}'+\vec{k}'')/2,
              (\vec{k}'-\vec{k}'')/2,
              \vec{k}'+\vec{k}'';
              \epsilon(\vec{k}')+\epsilon(\vec{k}''))$
and implicitly also depends on $P[\phi^*,\phi;t]$, because it depends on the average occupation numbers 
$N(\vec{k};t) \equiv {\rm Tr}[\rho(t_0) 
                           \psi^{\dagger}_{\vec{k}}(t)
                           \psi_{\vec{k}}(t)]$. 
The same is in fact true for the various selfenergies as we will see now in more detail. 

In the above Fokker-Planck equation the nonlinear terms proportional to $V^{(\pm),K}(\vec{k},\vec{k}',\vec{k}'')$ should be regarded as giving corrections to the linear terms proportional to the selfenergies $\hbar\Sigma^{(\pm),K}(\vec{k};t)$ due to the condensate. The effects of the above condensate particles have already been included correctly in the linear terms and should not be accounted for twice. Therefore, in the normal state of the gas we must use the linearized version of Eq.~(\ref{FP}). Restricting ourself to this case first, we see that the real part of $\hbar\Sigma^{(\pm)}(\vec{k};t)$, which we call $S(\vec{k};t)$ in the following, describes the change in the instantaneous energy levels of the gas due to the interactions. Indeed, the instantaneous energy levels obey 
$\hbar\omega(\vec{k};t) = \epsilon(\vec{k}) + S(\vec{k};t)$.  
From our experience with the semiclassical theory we know that the selfenergies change on a very short time scale of $O(\hbar/k_BT_c)$ for conditions near the critical temperature $T_c$. However, the change of the occupation numbers, as derived from the imaginary part $R(\vec{k};t)$ of $\hbar\Sigma^{(+)}(\vec{k};t)$, occurs only on a much longer time scale of $O((\Lambda_c/a)^2\hbar/k_BT_c)$ determined by the mean free path of the atoms. As a result it is very accurate to apply an adiabatic approximation and use 
\end{multicols}
\begin{equation}
\label{S}
S(\vec{k};t) \simeq 2nT^{(+)}(\vec{0},\vec{0};0) -
   2 \int \frac{d\vec{k}'}{(2\pi)^3} 
     \int \frac{d\vec{k}''}{(2\pi)^3}~
       \left| V^{(+)}(\vec{0},\vec{k}',\vec{k}'') \right|^2 
          N(\vec{k}';t) N(\vec{k}'';t) 
          \frac{1-\cos(\hbar \vec{k}' \cdot \vec{k}''(t-t_0)/m)}
                {\hbar^2 \vec{k}' \cdot \vec{k}''/m}~,
\end{equation}
where $T^{(+)}(\vec{0},\vec{0};0) = 4\pi a\hbar^2/m$ is the two-body T-matrix at zero momenta and energy, and
\begin{eqnarray}
R(\vec{k};t) = - 2\pi
     \int \frac{d\vec{k}'}{(2\pi)^3} 
     \int \frac{d\vec{k}''}{(2\pi)^3}~ 
       \left| V^{(+)}(\vec{k},\vec{k}',\vec{k}'') \right|^2 
          \delta\left(\epsilon(\vec{k}') + \epsilon(\vec{k}'') -
                 \epsilon(\vec{k}' + \vec{k}''-\vec{k}) -
                 \epsilon(\vec{k}) \right)
                                \hspace*{0.7in}  \nonumber \\
   \times \left[ (1 + N(\vec{k}';t))(1 + N(\vec{k}'';t))
                      N(\vec{k}' + \vec{k}''-\vec{k};t) -
                  N(\vec{k}';t) N(\vec{k}'';t)
                      (1 + N(\vec{k}' + \vec{k}''-\vec{k};t))
          \right]~,
\end{eqnarray}
as also expected from a Fermi's Golden Rule calculation of 
$\partial N(\vec{k};t)/ \partial t$. Moreover, in the same aproximation 
\begin{eqnarray}
\hbar\Sigma^K(\vec{k};t) = - 4\pi i
     \int \frac{d\vec{k}'}{(2\pi)^3} 
     \int \frac{d\vec{k}''}{(2\pi)^3}~ 
       \left| V^{(+)}(\vec{k},\vec{k}',\vec{k}'') \right|^2 
          \delta\left(\epsilon(\vec{k}') + \epsilon(\vec{k}'') -
                 \epsilon(\vec{k}' + \vec{k}''-\vec{k}) -
                 \epsilon(\vec{k}) \right)
                                \hspace*{0.5in}  \nonumber \\
   \times \left[ (1 + N(\vec{k}';t))(1 + N(\vec{k}'';t))
                      N(\vec{k}' + \vec{k}''-\vec{k};t) +
                  N(\vec{k}';t) N(\vec{k}'';t)
                      (1 + N(\vec{k}' + \vec{k}''-\vec{k};t))
          \right]~. 
\end{eqnarray}
\vspace*{-0.2in}
\begin{multicols}{2}
\noindent
Notice that in equilibrium the above relations imply that 
$\hbar\Sigma^K(\vec{k}) = 2i (1 + 2N(\vec{k})) R(\vec{k})$,
which is just the famous fluctuation-dissipation theorem.

In the normal state the Fokker-Planck equation in Eq.~(\ref{FP}) thus combines the kinetic evolution of the occupation numbers due to the quantum Boltzmann equation with the coherent evolution of the energy levels given by 
$\hbar\omega(\vec{k};t) \simeq \epsilon(\vec{k}) + S(\vec{0};t)$. In particular, the solution of the Fokker-Planck equation in the limit $t \rightarrow \infty$ is due to the fluctuation-dissipation theorem correctly given by
\begin{eqnarray}
\label{equi}
P[\phi^*,\phi;\infty] = 
  \prod_{\vec{k}} \alpha(\vec{k}) 
     \exp \left\{ -\alpha(\vec{k}) 
                  \left| \phi_{\vec{k}} \right|^2
          \right\}~,   \nonumber
\end{eqnarray}
with $\alpha(\vec{k}) = 1/(N(\vec{k})+1/2)$, $N(\vec{k})$ the Bose distribution evaluated at 
$\epsilon(\vec{k}) + S(\vec{0};\infty) - \mu_0$ and $S(\vec{0};\infty)$ determined selfconsistently by
Eq.~(\ref{S}). In equilibrium, Bose-Einstein condensation therefore occurs at a temperature such that $S(\vec{0};\infty) = \mu_0$. From the semiclassical analysis we know that this can only be fulfilled in the case that $a>0$ and then leads to a critical temperature $T_c$ for the interacting gas that is higher than that for the noninteracting gas. Interestingly, the same conclusion was recently also reached by a completely different calculation \cite{michel}.  

Returning to the nonequilibrium problem of interest, we see that if the gas is quenched sufficiently far into the critical region, the effective chemical potential 
$\mu_{cl}(t) \equiv \mu_0 - S(\vec{0};t)$ will become positive during the kinetic evolution of the gas towards equilibrium. At that time, say at $t=t_1$, the gas is unstable against the formation of a condensate. To determine how the condensate actually grows after the instability has occured, we must return to our full nonlinear Fokker-Planck equation and discuss the coherent stage that now follows. Neglecting therefore the kinetic, i.e.\ imaginary, terms in the right-hand side of Eq.~(\ref{FP}) and making use of the fact that in the coherent stage of the evolution we are mainly interested in momenta 
$\hbar k < \hbar\sqrt{na}$, we can easily show that a solution to the Fokker-Planck equation is  
\begin{eqnarray}
P[\phi^*,\phi;t] =
  \int d[\phi^*_1]d[\phi_1]~P[|\phi_1|^2;t_1]
       \delta[|\phi|^2 - |\phi_{cl}|^2]~,  \nonumber 
\end{eqnarray}
where $P[|\phi_1|^2;t_1]$ is the probability distribution after the kinetic stage and $\phi_{cl}(\vec{x},t)$ is the solution of the nonlinear Schr\"odinger equation
\begin{eqnarray}
i\hbar \frac{\partial \phi_{cl}}{\partial t} =
    \left\{ - \frac{\hbar^2 \nabla^2}{2m} - \mu_{cl}(t) 
            + T^{(+)}(\vec{0},\vec{0},\vec{0};0) |\phi_{cl}|^2 
    \right\} \phi_{cl}                                \nonumber
\end{eqnarray} 
with the initial condition 
$\phi_{cl}(\vec{x},t_1) = \phi_1(\vec{x})$.

Clearly, with this approximate solution to the Fokker-Planck equation we have essentially rederived the semiclassical theory of the coherent stage. It is interesting to note that in contrast with the quantum theory for the turning on of a laser \cite{mandel}, this coherent stage of the evolution is important. The reason for this is that in the case of a laser there is only one mode for the photon field and the nonlinear Schr\"odinger equation then only affects the phase of this mode, with the result that $P[\phi^*,\phi;t] = P[|\phi|^2;t_1]$. However, in a Bose gas we are dealing with many modes and the nonlinear Schr\"odinger equation actually leads to the building up of population in the one-particle ground state due to a depopulation of the lowest excited states. Introducing the phase $\chi_{cl}$ of $\int d\vec{x}~\phi_{cl}$, the latter picture arises roughly speaking because the instantaneous energies of the excited states are 
$\hbar\omega(\vec{k};t) = 
   [(\epsilon(\vec{k}) + \hbar \dot{\chi}_{cl} 
                       - \mu_{cl} + 2n_{c}T^{(+)})^2
                                   - (n_c T^{(+)})^2]^{1/2}$,
which are imaginary for low momenta since the time derivative
$\hbar \dot{\chi}_{cl}(t) < \mu_{cl}(t) 
                   - n_c(t) T^{(+)}(\vec{0},\vec{0},\vec{0};0)$
as the population of the lowest excited states decreases in favor of the condensate density $n_c(t)$.

Having formed a small condensate density in this way, the gas then has to evolve towards an equilibrium between the condensate  and the noncondensed part of the gas. For this the imaginary or kinetic terms in the right-hand side of Eq.~(\ref{FP}) are responsible. Indeed, using the explicit expressions for $V^{(\pm),K}(\vec{k},\vec{k}',\vec{k}'')$ following from Eq.~(\ref{LS}), we find that in lowest order the presence of the condensate has the effect of replacing $N(\vec{k}'+\vec{k}''-\vec{k};t)$ by $N(\vec{k}'+\vec{k}''-\vec{k};t) 
  + V n_c(t) (2\pi)^3\delta(\vec{k}'+\vec{k}''-\vec{k})$
in the normal state expressions for $R(\vec{k};t)$ and $\hbar\Sigma^K(\vec{k};t)$. The kinetic equations for the superfluid state of the gas are therefore essentially the same as those recently studied by Semikoz and Tkachev \cite{kinetics}. Although these do not take properly account of the coherence effects at low momenta $\hbar k < \hbar\sqrt{na}$, they are sufficiently accurate for our purposes because for the quantum gases of interest we always have that 
$\hbar/\Lambda \gg \hbar\sqrt{na}$ near the critical temperature.
      
This completes our discussion of the quantum theory for the formation of a condensate in a weakly-interacting Bose gas. Summarizing, we have derived a Fokker-Planck equation for the probability distribution of the order parameter 
$\langle \psi(\vec{x},t) \rangle$, which gives a desciption of both the kinetic and coherent stages of Bose-Einstein condensation. This Fokker-Planck equation implies that the order parameter obeys a nonlinear Schr\"odinger equation with a Langevin noise term $\eta(\vec{x},t)$, which in a good approximation has a Gaussian probability distribution that fulfills the fluctuation-dissipation theorem. We have also discussed the qualitative behavior of the solutions to the Fokker-Planck equation. We hope to report on a more quantitative numerical study in the near future.  

It is a pleasure to thank Keith Burnett for many stimulating discussions and for making possible the visit to Oxford, that initiated the above. I also thank Crispin Gardiner and Steve Girvin for illuminating comments.

\end{multicols}
\end{document}